\documentclass[preprint,preprintnumbers,amsmath,amssymb]{revtex4}

\usepackage{graphicx}
\usepackage{dcolumn}
\usepackage{bm}
\begin{document}

\title{A method for measuring the nonlinear response in dielectric spectroscopy through third harmonics detection}

\author{C. Thibierge}
\author{D. L'H\^ote}
\email{denis.lhote@cea.fr}
\author{F. Ladieu}
\email{francois.ladieu@cea.fr}
\author{R. Tourbot}

\affiliation{Service de Physique de l'Etat Condens\'e (CNRS/MIPPU/URA 2464),
DSM/IRAMIS/SPEC CEA Saclay, Bat.772, F-91191 Gif-sur-Yvette Cedex France}



\vskip 2cm
\begin{abstract}
We present a high sensitivity method allowing the measurement of the non linear dielectric susceptibility of an insulating material at finite frequency. It has been developped for the study of dynamic heterogeneities in supercooled liquids using dielectric spectroscopy at frequencies 0.05 Hz$\leq$ $f$ $\leq$ $3 \times 10^4$ Hz . It relies on the measurement of the third harmonics component of the current flowing out of a capacitor. We first show that standard laboratory electronics (amplifiers and voltage sources) nonlinearities lead to limits on the third harmonics measurements that preclude reaching the level needed by our physical goal, a ratio of the third harmonics to the fundamental signal about $10^{-7}$. We show that reaching such a sensitivity needs a method able to get rid of the nonlinear contributions both of the measuring device (lock-in amplifier) and of the excitation voltage source. A bridge using two sources fulfills only the first of these two requirements, but allows to measure the nonlinearities of the sources. Our final method is based on a bridge with two plane capacitors characterized by different dielectric layer thicknesses. It gets rid of the source and amplifier nonlinearities because in spite of a strong frequency dependence of the capacitors impedance, it is equilibrated at any frequency. We present the first measurements of the physical nonlinear response using our method. Two extensions of the method are suggested.

\end{abstract}

\maketitle

\section{\label{Introduction} Introduction}

Measuring the nonlinear response of a physical system to an excitation is a way to investigate physical properties often unreachable through the linear response. Understanding nonlinear effects allowed decisive breakthrough in condensed matter physics. Spin glasses\cite{Dzya78,Levy86,Levy88,Jon98,Jon07,Nai07}, ferroelectric, freezing, or dipolar glass transitions\cite{Fay67,Mag86,Ike87,Gar01,Hem95,Hemb96,Hem96,Lev98,Kle03,Bob03,Vod04,Dec06,Mig07}, isotropic-liquid crystal transition \cite{Dro05,Dro02} or binary mixtures \cite{Rzo93,Gor96}, superconductivity\cite{Cla91,Lee93,Sen05,Bau06,Kuz07,Suz07}, field\cite{Midd94,Li99,Ladi00,God06,Ham07} or heating\cite{Ros65,Ahlu82,Bir97,Letu03,Gur06,Gon07} effects in electrical transport, heating due to electric field excitation of supercooled liquids\cite{Rich06,Rich07} are a few among many topics where non linear measurements have proven to be a precious tool.

The detection of harmonics of the fundamental response is a powerful method for studying nonlinear effects. It allows to get rid of the linear response signal, which usually is much larger than the sought nonlinear signals. 
We consider in this paper the nonlinear response of a dielectric system to a time-dependent electric field $E(t)$. The method we present allows to extract very low level harmonics in the response to a sinusoidal excitation.
The most general relationship relating the response (polarisation $P(t)$) to the excitation $E(t)$ can be written as a series expansion in $E$ (the even terms are forbidden because of the symmetry with respect to field reversal $E(t) \rightarrow -E(t)$):
\begin{equation}
\frac {P(t)}{\epsilon_0} = \int_{-\infty}^{\infty}\chi_1(t-t')E(t')dt' + \iiint_{-\infty}^{\infty}\chi_3(t-t'_1,t-t'_2,t-t'_3) 
\times E(t'_1)E(t'_2)E(t'_3)dt'_1dt'_2dt'_3 + ...
\label{eqchidet}.
\end{equation}
In this equation $\epsilon_0$ is the dielectric constant of vacuum, $\chi_1$ the linear susceptibility and $\chi_3$ the cubic nonlinear susceptibility. The dots in Eq.~\ref{eqchidet} indicate an infinite sum involving higher order non linear susceptibilities $\chi_5$, etc. Note that causality implies $\chi_{i}(t<0) =0$. The Fourier transform of Eq.~\ref{eqchidet} for a purely a.c. field $E=E_0\cos(\omega t)$ gives  
\begin{eqnarray}
\frac {P(\omega')}{\epsilon_0} &=& \frac{E_0}{2}\left[ \chi_1(\omega)+ \frac{3E_0^2}{4}\chi_3(-\omega,\omega,\omega)+... \right] \delta(\omega'-\omega) \nonumber \\
&+& \frac{E_0}{2}\left[ \chi_1(-\omega)+ \frac{3E_0^2}{4}\chi_3(\omega,-\omega,-\omega)+... \right] \delta(\omega'+\omega) \nonumber \\
&+& \frac{E_0^3}{8}\chi_3(\omega,\omega,\omega) \delta(\omega'-3\omega) \nonumber \\
&+& \frac{E_0^3}{8}\chi_3(-\omega,-\omega,-\omega) \delta(\omega'+3\omega)+... 
\label{eqchideomega} ,
\end{eqnarray}
where the polarization $P$ and the susceptibilities $\chi_i$ are now in the frequency domain and the dots indicate again infinite sums involving higher order terms.
The response $P(t)$ to $E(t)$ = $E_0 \cos(\omega t)$ can thus be written 
\begin{equation}
P(t)/\epsilon_0 = Re \left[(E_0 \chi_1(\omega)+3/4E_0^3 \chi_{\bar{3}}(\omega)+ ...) e^{-i\omega t}\right] 
+ Re \left[1/4E_0^3 \chi_3(\omega) e^{-i3\omega t} + ...\right]+ ...,               
\label{eqchideomegapract1} 
\end{equation}
where we have used the fact that because $\chi_1$ and $\chi_3$ are real in the time domain, their Fourier transform verify 
$\chi_1^*(\omega)$ = $\chi_1(-\omega)$ and $\chi_3^*(\omega_1,\omega_2,\omega_3)$ = $\chi_3(-\omega_1,-\omega_1,-\omega_1)$ (the star denotes the complex conjugate), and the invariance of $\chi_3$ by permutation of its arguments. For simplicity, we write $\chi_3(\omega)$ = $\chi_3(\omega,\omega,\omega)$ and $\chi_{\bar{3}}(\omega)$ = $\chi_3(-\omega,\omega,\omega)$. Eq.~\ref{eqchideomegapract1} can be written 
\begin{eqnarray}
P(t)/\epsilon_0 = E_0(\chi'_1 \cos\omega t + \chi''_1 \sin\omega t) + 3/4 E_0^3(\chi'_{\bar{3}} \cos\omega t & \nonumber \\ 
+ \chi''_{\bar{3}} \sin\omega t)+... + 1/4 E_0^3(\chi'_3 \cos3\omega t + \chi''_3 \sin3\omega t)+...,\hfill &\nonumber \\
\label{eqchideomegapract2} 
\end{eqnarray}
where the susceptibilities $\chi_i$ are written as a function of their real and imaginary parts $\chi_i'$ and $\chi_i''$. For practical applications, the modulii and arguments $\left|\chi_i\right|$ and $\delta_i$ are rather used:
\begin{multline}               
P(t)/\epsilon_0 = E_0\left|\chi_1\right| \cos(\omega t - \delta_1) + 3/4 E_0^3\left|\chi_{\bar{3}}\right| \cos(\omega t - \delta_{\bar{3}}) + \\ 
+... + 1/4E_0^3\left|\chi_3\right| \cos(3\omega t - \delta_3)+...  
\label{eqchideomegapract3} 
\end{multline}

We see in the first and second terms of the rhs in Eqs~\ref{eqchideomegapract1}-\ref{eqchideomegapract3} that the nonlinear susceptibility $\chi_{\bar{3}}$ could be extracted from a measurement at the fundamental frequency by varying $E_0$. However, in most experiments $\left|\chi_1\right| \gg E_0^2 \left|\chi_{\bar{3}}\right|$, thus the nonlinear part can hardly be separated from the much larger $\chi_1$ linear term. On the contrary measuring the harmonics yields directly the physical information contained in $\chi_3$, $\chi_5$, etc. In our case the physical information of interest is contained in $\chi_3$ (see section~\ref{Motivation}). We thus chose to measure the third harmonics. However, its relative magnitude with respect to the fundamental was so low that we had to develop a special method for obtaining an accurate measurement.

Experimentally, various methods have been used to extract a very small third, fifth, etc. harmonics signal \cite{Gul65,Ros65,Fay67,Dixo88,Jun92,Lee93,Meno96,Bir97,Li99,Gar01}. A bridge technique has been often used: The first arm of the bridge contains the sample under study, and the second arm a well known impedance with zero nonlinear response\cite{Gul65}. This is the case for specific heat spectroscopy based on thermal diffusion into a thick sample from a thin metallic film that serves simultaneously as heater and thermometer\cite{Dixo88,Meno96,Bir97} or similarly for the study of a heating resistor where the third harmonics is related to its electrothermal parameters\cite{Ros65,Jun92}. The bridge technique has also been used to measure the nonlinear inductive response of a superconducting film\cite{Lee93}, the nonlinear behaviour of power piezoceramic materials\cite{Gar01}, etc. The bridge techniques evoked above cannot be used for our purpose because the balancing of the bridge at the fundamental frequency would be lost at the third harmonics: The strong frequency dependence of our sample impedance can hardly be mimicked by a combination of resistors and capacitors in the second arm. As a result, the third harmonics generated by the voltage source would not be cancelled in the bridge.
  
In this paper, we report on a two samples bridge method allowing to measure properly the third harmonics (i.e. the third term in the rhs of Eqs~\ref{eqchideomegapract1}-\ref{eqchideomegapract3}), and eventually other harmonics of the polarization induced by the ac $E$-field excitation of a dielectric sample placed between the two electrodes of a plane capacitor. This is done through the measurement of the third harmonics of the current induced by the ac voltage excitation. Such a measurement yields directly the cubic nonlinear term $\chi_3$ of the response (assuming that the higher order contributions related to $\chi_5$, $\chi_7$, etc. to the third harmonics are weak which is the case in our experiment). We developped it for studying supercooled liquids such as glycerol.  In part~\ref{Motivation} we first briefly summarize the physical interest of such measurements. Part ~\ref{Setup} is devoted to describing the experimental problems to solve. In part ~\ref{New setup} we present the two sample capacitors bridge method which allows to measure $\chi_3$. Part~\ref{Final results} gives our first measurements using this method. Finally, in part~\ref{Extensions} we consider possible developments or extensions of this method.

\section{\label{Motivation} Physical motivation: dynamical correlations in glass formers}

The physics of structural glasses still lacks a firm experimental basis for a growing length scale when the temperature $T$ decreases towards the glass transition temperature $T_g$ \cite{Edig00,Lun00,Dont01,Rich02,Ber05,Dal07}. A basic feature of glassforming supercooled liquids is the spectacular increase of the characteristic relaxation time $\tau_{\alpha}(T)$ as $T$ decreases towards $T_g$. $T_g$ is conventionally defined by $\tau_{\alpha}(T_g)$ $\simeq$ 100 s. $\tau_{\alpha}$ is often obtained from dielectric spectroscopy\cite{Lun00,Rich02,Ladi07}. Fig.~\ref{figchideomega} shows an example of linear dielectric susceptibility $\chi_1(\omega)$ measurement for glycerol (C$_3$O$_3$H$_8$, $T_g$ $\simeq$ 190 K) at $T$ = 211.8 K. The imaginary part $Im(\chi_1)$ is maximum, and the real part $Re(\chi_1)$ is approximately half its maximum plateau for $\omega_{\alpha} = 2\pi/\tau_{\alpha}$. At present, the fast decrease of $\omega_{\alpha}$ with $T$ has not received a unique microscopic physical interpretation, but a seminal concept \cite{Adam65,Whit04} is that of cooperative effects: When $T_g$ is approached, molecules belonging to larger and larger regions (called dynamical heterogeneities, DH) should move in a correlated way to allow relaxation. Correlation lengths probing such cooperative effects have been extracted experimentally, leading to length scales estimates of $4-12$ molecular diameters \cite{Trac98,Week00,Vida00,Edig00,Desc01,Sill02,Sinn05}. However these experiments were not able to test the expected increase of the correlation length as $T$ decreases towards $T_g$ which is of fundamental interest. For spin glasses the increase of the correlation length close to the critical temperature is associated to the divergence of $\chi_{i}$ ($i \ge 3$)\cite{Dzya78,Levy86,Levy88,Jon98,Jon07,Nai07}.
This suggests to investigate $\chi_{i}(T)$ ($i \ge 3$) close to the structural glass transition, although no divergence is expected. Up to very recently, the nonlinear susceptibility $\chi_3(\omega)$ had never been measured in these systems (see however \cite{Wu91,Rich06,Rich07}) contrary to spin glasses. However, some important theoretical progress was made recently by Bouchaud and Biroli \cite{Bouc05} who established that:

\begin{equation}
\chi_3(\omega) = \frac{\epsilon_{0} \chi_1^2(0)a^3}{k_B T} N^*_{corr}\cal{H}(\omega \tau_{\alpha}),
\label{eqchi3omega}
\end{equation}
where $a^3$ is the molecular volume and $\cal{H}$ is a complex scaling function which should have a maximum (in modulus) for $\omega\tau_\alpha \simeq 2\pi$. $N^*_{corr}$ is the maximum of the number of correlated particles $N_{corr}(t)$ which increases with the time $t$, reaches its maximum $N_{corr}^*$ for $t \sim \tau_{\alpha}$ and then goes to 0 for $t \rightarrow \infty$. This is reflected in the frequency space by the fact that $\left|\cal{H}(\omega \tau_{\alpha})\right|$ should be maximum for $\omega \sim \omega_{\alpha}$. Thus, measuring the nonlinear dielectric response $\chi_3$ gives the size $N^*_{corr}$ of the correlated regions and its $T$-dependence. 

What is the value of $\chi_3(\omega)$ that can be expected from Eq.~\ref{eqchi3omega} ? The answer cannot be very precise since little is known about $\cal{H}$ which, theoretically, should reach a maximum of ``order $1$'' for $\omega \tau$ ``of order $1$''\cite{Bouc05}. 
A conservative estimate of $\left| \chi_3 \right|_{max}$ = $max_{\omega}(\left|\chi_3(\omega)\right|)$ for $N_{corr}^* = 1$ can be drawn by assuming $max_{\omega}(\left|\cal{H}(\omega)\right|)$ = 1. We consider the case of glycerol at 200 K. $\chi_1(0)$ should be replaced by $\chi_1(0) - \chi_1(\infty)$ = $\Delta \epsilon \approx 72$ ($\chi_1(\omega) = \epsilon(\omega) - 1$) because only the contribution of the molecular motion to the dielectric response is considered. $a^3$ $\simeq$ 0.115 nm$^3$ is obtained from the density at 200 K \cite{Rei05} and the molecular mass 92.09 g. As a reference, we take $\left|\epsilon(\omega_\alpha = 2\pi/\tau_\alpha)\right| = \left| 1+ \chi_1(\omega_\alpha)\right|$ $\simeq$ 52, and find by using Eq.~\ref{eqchi3omega}, $\left| \chi_3 \right|_{max}/\left|\epsilon(\omega_{\alpha})\right|$ $\simeq$ $ 3.685\times 10^{-17}$ m$^2$.V$^{-2}$. 
We shall use this value in section~\ref{Setup a} to determine the sensitivity required in our experiments.

\section{\label{Setup} Experimental setup and requirements for the electronics}

\subsection{\label{Setup a} Cryogenics and nonlinear dielectric measurements}

The experiments were performed in a cryostat connected to a cryogenerator with a base temperature of 10 K. The experimental cell is a closed metallic box placed in vacuum, related to the cold stage of the cryogenerator through a thermal impedance \cite{Ladi07}. The temperature $T$ in the cell is set by a PID LakeShore$^{\hbox{\textregistered}}$
331 controller which regulates the heating power flowing through the thermal impedance. The cell contains two independent plane capacitors of equal surfaces $S$ but different thicknesses $L_{thick}$ and $L_{thin}$ (The reason for two capacitors instead of one is given in section ~\ref{New setup}). The capacitors are used to measure the dielectric susceptibilities of a glassforming liquid (glycerol) placed between the electrodes. For each capacitor, the two electrodes are immersed in the liquid. We used two experimental setups A and B. In setup A, the electrodes of the two capacitors are polished and gold plated brass squares ($S \simeq 5.5$ cm$^2$), separated by three $0.03$ cm$^2$ Mylar$^{\hbox{\textregistered}}$ discs of thickness $L_{thin} \simeq 30\ \mu$m for the thinner sample and $L_{thick} \simeq 60\ \mu$m for the thickest sample. Setup B is an improved version, in which the electrodes are gold plated copper disks (metallic mirrors) with $L_{thin} \simeq 19\ \mu$m, $L_{thick} \simeq 41\ \mu$m and $S$ = 3.14 cm$^2$. We minimized the spacers volume in order to minimize their contibution to the dielectric response of the capacitors. The supercooled liquid was allowed to flow in and out of the volume between the electrodes in order to avoid pressure effects due to the different dilatation coefficients of the supercooled liquid and the spacers. The results presented in sections~\ref{Setup},~\ref{New setup} (resp.~\ref{Final results}) were obtained using setup A (resp. B). A pressure of 2 bars of Argon was set in the cell at room temperature to ensure that, at the working temperature of $\sim$ 200 K, the pressure in the cell remains above $1$ bar, preventing the formation of bubbles of the gas adsorbed at the surface of the electrodes or dissolved in glycerol. Coaxial shielding was ensured all along the circuit, from the measurement apparatus at room temperature down to the experimental cell at low temperature.

When a voltage $V(t)=V_0\cos(\omega t)$ is applied to the electrodes of a plane capacitor of thickness 
$L$ and surface $S$, the resulting field $E(t)=V(t)/L$ induces a polarization $P(t)$ and an electrical displacement $D(t)=\epsilon_0 E(t) + P(t)$. As a result, a current $I(t)$ flows in the circuit, the current density $I/S$ being the time derivative of $D(t)$. Using Eq~\ref{eqchideomegapract1}, in which only the first terms of the fundamental and third harmonics series are kept, we thus have
\begin{equation} 
I(t)=Re(I(\omega)e^{-i\omega t} + I(3\omega)e^{-i3\omega t}),
\label{eqI1I3}
\end{equation}
where $I(\omega)$ and $I(3\omega)$ are complex numbers giving the magnitude and the phase of the two components of the current. The linear part of the current is given by $I(\omega)=Y(\omega)V_0$ where the admittance $Y(\omega)= G + i \omega C$ comprises the conductance $G=\epsilon_0 \omega Im(\chi_1(\omega)) S/L$ and the capacitance $C=\epsilon_0 Re( 1+ \chi_1(\omega)) S/L$. As depicted in the upper inset of Fig.~\ref{figchideomega}, we measure the current $I(\omega)$ through the voltage drop $V_A(\omega)$ across a resistor $r$ = 1 k$\Omega$ in series with the capacitor. All the voltage measurements presented in this paper were performed with a standard commercial phase sensitive lock-in amplifier with an input impedance $Z_L$ of 10 M$\Omega$ in parallel with 25 pF. The accuracy of the $V_A(\omega)$ measurement is typically not much better than $0.1\%$, which justifies that in the analysis of $I(\omega)$ we neglect the contributions of the nonlinear terms of the series giving the response at the fundamental frequency. The voltage sources were limited to $V_s(\omega)$ $\leq$ 7 V (rms) for the measurements with setup A presented in sections~\ref{Setup},~\ref{New setup} and~\ref{results} and to $V_s(\omega)$ $\leq$ 14 V (rms) for those with setup B presented in section~\ref{Final results}. The frequency range was about 0.05 Hz$\leq$ $\omega/2\pi$ $\leq$ $3 \times 10^4$ Hz. All the voltage and current magnitudes presented hereafter are rms.

The calculation of the third harmonics component of the current, $I(3\omega)e^{i3\omega t}$ (see Eq.~\ref{eqI1I3}) from the time derivative of $D$ (neglecting the contribution of the terms proportional to $\chi_5$, $\chi_7$,...), gives  
\begin{equation} 
I(3\omega) = \frac{-3i}{4} \epsilon_0 \omega \chi_3(\omega)S \left( \frac{V_{0}}{L} \right)^3.
\label{eqI3}
\end{equation}
In what follows, we investigate the measurement of $I(3\omega)$ (and $I(\omega)$) for our liquid dielectric capacitors (that will be called ``sample capacitors'' or ``samples'' for simplicity).  
As the third harmonics current is simply added to the fundamental (Eq.~\ref{eqI1I3}),  the sample can be represented for what concerns the third harmonics by a current source given by Eq.~\ref{eqI3} and placed in parallel with the liquid dielectric capacitor, whose complex admittance is $Y(3\omega)$ (see lower inset of Fig.~\ref{figchideomega}). Note that due to the Thevenin theorem, one could, as well, represent the sample at $3 \omega$ by a voltage source $V_{sample}= I(3\omega)\times Z(3\omega)$ in series with the impedance $Z(3\omega)=1/Y(3\omega)$.

From the estimate of $\left|\chi_3\right|$ in section~\ref{Motivation}, we can determine the required sensitivity of our measurements. The complex currents $I(\omega)$ and $I(3\omega)$ being obtained from the complex fundamental and third harmonics components of the displacement, we have $I(\omega)=i\omega S (1+\chi_1)\epsilon_0 E_0$ and $I(3\omega)= 3i/4\omega S \chi_3 \epsilon_0 E_0^3 $, thus   
\begin{equation}
\frac{\left|I(3\omega)\right|}{\left|I(\omega)\right|} = \frac{3\left|\chi_3(\omega)\right|}{4\left|1+\chi_1(\omega)\right|}E_0^2
\simeq 2.76\times 10^{-17}E_0^2,
\label{eqI3surI1}
\end{equation}
where the numerical factor is obtained for $\omega = \omega_\alpha$, and $E_0$ is in V/m. We have assumed, according to the theoretical prediction~\cite{Bouc05} that $\left|\chi_3(\omega_{\alpha})\right|$ $\simeq$ $\left|\chi_3\right|_{max}$, and used the value of $\left|\chi_3\right|_{max}$ estimated in section~\ref{Motivation}.
The maximum field $E_0$ in our experiments is 200 kV/m for setup A (7 V on 30 $\mu$m), and 740 kV/m for setup B (14 V on 19 $\mu$m), thus $\left|I(3\omega) /I(\omega)\right|$ $\simeq$ $1.1\times 10^{-6}$ or $1.5\times 10^{-5}$.
The required sensitivity on the measurement of $\left|I(3\omega)/I(\omega)\right|$ is lower than these values, about $1\times 10^{-7}$ because of the uncertainty on $\cal{H}$ and to allow the measurement of $\chi_3(\omega)$ when its magnitude is below $\left|\chi_3\right|_{max}$. We thus look for a setup whose relative parasitic contributions at $3\omega$ remain below $10^{-7}$.

\subsection{\label{Setup b} Nonlinear behavior of the lock-in amplifier}
In this section we see why the simplest possible circuit, depicted in the upper inset of Fig.~\ref{figchideomega}, cannot work to detect $I(3\omega)$. For simplicity we assume that the resistance $r$ is small with respect both to the impedance of the sample $\left|Z(\omega)\right|$ and to the input impedance of the lock-in amplifier $\left|Z_L(\omega)\right|$. The voltage drop accross $r$ at the fundamental frequency is thus $V_A(\omega) \simeq V_s r/Z(\omega)$. Fig.~\ref{figlockin} shows that  applying a $1\omega$ signal directly at the input of the lock-in amplifier induces the measurement of a rather important $3\omega$ signal. We found by using the method described in section~\ref{Setup c} that this signal does not come from the source (which is that of the lock-in amplifier) which is expected to generate small harmonics in addition to the $1\omega$ signal. It thus comes from the nonlinearities of the lock-in amplifier itself. We checked that the behavior reported in Fig.~\ref{figlockin} does depend neither on the frequency, nor on the output impedance of the source $r_0$.

The magnitude of the $3\omega$ signal due to the lock-in amplifier nonlinearities is much larger than the one expected from a glycerol sample in the simplest possible circuit, i.e. the sample in series with a measuring resistance $r$, see upper inset of Fig.~\ref{figchideomega}. Assuming $r \ll \vert Z(\omega) \vert$, one finds that $I(3\omega)$, see Eq.~\ref{eqI3}, yields a voltage $V_{A,sample}(3\omega) \simeq r I(3\omega)$, while $I(\omega) \simeq V_s/Z(\omega)$, thus:
\begin{equation}
 \frac{V_{A,sample}(3\omega)}{V_s} \simeq \frac{r}{Z(\omega)}\frac{I(3\omega)}{I(\omega)}
\label{eqsetupb}
\end{equation} 

We have seen in section~\ref{Setup a} that a typical maximum expected value of $\left|I(3\omega)/I(\omega)\right|$ was about 10$^{-6}$ - 10$^{-5}$. There is an optimum of the ratio $r/\left| Z(\omega) \right|$ which maximizes $\left|V_{A,sample}(3\omega)\right|$ (see section~\ref{New setup1}), and a reasonnable value is 0.1. Thus Eq.~\ref{eqsetupb} gives $V_{A,sample}(3\omega)/V_s$ $\approx$ $10^{-7}$ - $10^{-6}$. The corresponding $V_s$ is of the order of 10 V and gives $V_A(\omega) \simeq$ 1 V, thus by using Fig.~\ref{figlockin} we find that the lockin nonlinearities give $V_A(3\omega)/V_s \ge 10^{-3}$, i.e. values much larger than the sample contribution.
 
To summarize this section, we found that the nonlinearities of a standard lock-in amplifier are such that, due to our very low physical $\left|I(3\omega)/I(\omega)\right|$ ratio, the $1\omega$ component must be ``removed from the signal'', before it enters into the lockin for the $3\omega$ detection. This is why, in the next sections, we shall report results obtained with bridges performing a ``$1\omega$ subtraction''.

\subsection{\label{Setup c} Nonlinear behavior of the voltage source}
\subsubsection{\label{pont2sources} Two sources bridge}
Let us consider the circuit depicted in the inset of Fig.~\ref{figpont2sources}. It uses two $1\omega$ voltage sources $V_{s,1}(\omega)$ and $V_{s,2}(\omega)$ with a common ground and a tunable relative phase shift. The lock-in amplifier (input impedance $Z_L$) is used to measure the voltage at point $A$ which is connected to the reference resistor $R$ and to the sample capacitor (impedance $Z(\omega)$). Considering, for simplicity, the case $R,\left|Z\right| \ll \left|Z_L\right|$, we calculate the voltage at point $A$ at the fundamental frequency :
\begin{equation}
V_A (\omega)= \frac{Z(\omega)V_{s,1}(\omega) + RV_{s,2}(\omega)}{Z(\omega)+R}.
\label{eqVApont2s}
\end{equation}
The balancing condition which fulfills $V_A(\omega)$ = 0 (see section ~\ref{Setup b}) is
\begin{equation}
\frac{V_{s,1}(\omega)}{V_{s,2}(\omega)} = -\frac{R}{Z(\omega)}.
\label{eqequilpont2s}
\end{equation}
We used a Tektronix$^{\hbox{\textregistered}}$ AFG3102 dual channel voltage source for which the relative phase between $V_{s,1}(\omega)$ and $V_{s,2}(\omega)$ is tunable with an accuracy of $0.01$ degree, while  $\left|V_{s,1}(\omega)\right|$ and $\left|V_{s,2}(\omega)\right|$ can be tuned with a relative precision of $10^{-4}$. As a consequence, $\left|V_A(\omega)/V_{s,2}(\omega)\right|$ can be made as small as $5\times 10^{-5}$. As a result, we find by extrapolating the curve of Fig.~\ref{figlockin}, that the lock-in nonlinearities contribution to $V_A(3\omega)$ remains much below 2 nV for the maximum voltage $V_{s,2}(\omega)=7$ V (which gives $\left|V_A(\omega)\right|$ $\simeq$ $5\times 10^{-5} \times 7$ V = $0.35\times 10^{-3}$ V). 

This value $\ll$ 2 nV is small in comparison with the expected physical signal $V_{A,sample}(3\omega)$. The latter can be estimated by using the circuit equivalent to the sample at $3\omega$: a current source $I(3\omega)$ in parallel with the sample of impedance $Z(3\omega)$ (see section~\ref{Setup a} and Fig.~\ref{figchideomega} ).
\begin{equation}
 \frac{V_{A,sample}(3\omega)}{V_{s,2}(\omega)} = \frac{R/Z(\omega)}{1+R/Z(3\omega)}\frac{I(3\omega)}{I(\omega)},
\label{eqsetupc}
\end{equation} 
where we have assumed $R,\left|Z \right|\ll \left|Z_L\right|$. We chose $R \simeq 0.084 \vert Z(\omega)\vert$ (The maximum voltage cannot be applied to the sample if $\vert R/Z(1\omega)\vert \ge 1$). For the capacitor we used ($L_{thin}$ = 30 $\mu$m), $\omega$ = $2\pi/\tau_{\alpha}$ (see section~\ref{Motivation}) and $V_{s,2}(\omega)$ = 7 V, and by using the value $\left|I(3\omega)/I(\omega)\right|$ $\simeq$ $10^{-7}$ obtained in section~\ref{Setup a} as the required sensitivity of our measurements, as well as $\left|Z(\omega_{\alpha})/Z(3\omega_{\alpha})\right|$ $\approx$ 1.85, we find $\left|{V_{A,sample}(3\omega)}\right|$ $\simeq$ $6\times10^{-8}$ V $>$ 2 nV . Thus the problem of the lock-in nonlinear contribution is solved by the two sources bridge.   
 
Despite this important feature, we shall see now that the two sources bridge does not reach the required resolution of $\left|I(3\omega)/I(\omega)\right|\simeq 1\times 10^{-7}$. This comes from the fact that the two sources, as any active electronic device, have a non-zero harmonic distortion: They generate $V_{s,1}(3\omega)$ and $V_{s,2}(3\omega)$ voltages in addition to $V_{s,1}(\omega)$ and $V_{s,2}(\omega)$. Thus a possible remaining problem is the contribution of $V_{s,1}(3\omega)$ and $V_{s,2}(3\omega)$ to $V_A(3\omega)$. To reduce it, we low-pass filtered the sources outputs with a dual channel six poles elliptic active filter, with a corner frequency $f_c$ chosen close to the working frequency $\omega/2\pi$. The filter damps the $3\omega$ component of the incoming signal, while it does not affect the $1\omega$ component. We verified that this filtering reduced significantly the harmonics magnitude. Fig.~\ref{figpont2sources} shows the measured $V_A(3\omega)$ for the two sources bridge at $\omega$ = $\omega_{\alpha}$. The results are the same for the thin and the thick sample of setup A ($L_{thin}$ = 30 $\mu$m and $L_{thick}$ = 60 $\mu$m), which shows that the measured signals do not come from the samples. Indeed, Eq.~\ref{eqI3} gives a current $\left|I(3\omega)\right|$ eight times larger for the thin sample than for the thick one. We took $R$ $\simeq$ $0.084 \left|Z(\omega)\right|$ in order to keep similar values of $V_{s,1}(\omega)$ and $V_{s,2}(\omega)$ for the thin and the thick sample measurements, thus, from Eq.~\ref{eqsetupc}, the expected ratio of $V_A(3\omega)$ for the two samples is 4, instead of the ``no change" result of Fig.~\ref{figpont2sources}. Thus we conclude that the third harmonics generated by the voltage sources are responsible for the results of Fig.~\ref{figpont2sources}. We verified that the expected physical value $V_{A,sample}(3\omega)$ (see section~\ref{Setup a}) is indeed lower than the measured $V_A(3\omega)$. 

A closer look at the equilibrium condition (Eq. ~\ref{eqequilpont2s}) of our two sources bridge reveals that once the circuit is balanced at $1 \omega$, it should not be in general balanced at $3\omega$, for two reasons. First, because of the strong change of $Z(\omega)$ when going from $1\omega$ to $3\omega$ (due to the $\chi_1(\omega)$ dependence, see Fig.~\ref{figchideomega}) while $R$ remains constant. Note that to mimick the $Z(\omega)$ dependence by using an impedance made of resistors and capacitors instead of $R$ would be hardly feasible because its components should be changed and tuned for each frequency and temperature. Second, and more importantly, because $V_{s,1}(3\omega)$ differs from $V_{s,2}(3\omega)$ as can be deduced from the comparison of the $3\omega$ signals measured when the two sources are exchanged (see Fig.~\ref{figpont2sources}).

\subsubsection{\label{Harmonics Source} Measurement of the harmonics signals generated by the sources}

The two sources bridge cannot be used to measure the nonlinear response of the samples, however it can be used to measure with a great accuracy the harmonics $V_{s,1}(n\omega)$ and $V_{s,2}(n\omega)$ of our voltage sources. This is of general interest for characterizing the harmonic distortion of any source, and it will be used in section ~\ref{New setup} (Note that we propose another method in section~\ref{Final results}). We now consider only the thickest sample of setup A for which $I(3\omega)$ is negligible. Assuming that the voltage measured at point $A$ at frequency $n\omega$,$V_A(n\omega)$, is due to the sources, we find for the circuit in the inset of Fig.~\ref{figpont2sources}
\begin{equation}
V_A(n \omega) = \frac{yV_{s,1}(n\omega)+Y(n\omega)V_{s,2}(n\omega)}{y+Y(n\omega)+Y_L(n\omega)},
\label{eqVAnomega}
\end{equation}
with $y=1/R$, $Y(n\omega)=1/Z(n\omega)$ and $Y_L(n\omega)=1/Z_L(n\omega)$. For any given $n$, another equation is necessary to determine the unknown complex quantities $V_{s,1}(n\omega)$ and $V_{s,2}(n\omega)$. It is obtained by exchanging the two sources in the circuit. The measured signal becomes $V_{A,ex.}$:
\begin{equation}
V_{A,ex.}(n \omega) = \frac{Y(n\omega)V_{s,1}(n\omega)+yV_{s,2}(n\omega)}{y+Y(n\omega)+Y_L(n\omega)}.
\label{eqVAprimenomega}
\end{equation}
Note that $V_{s,1}(n\omega)$ and $V_{s,2}(n\omega)$ are two functions of respectively $V_{s,1}(1 \omega)$ and $V_{s,2}(1 \omega)$ which can be different from each other. Thus, for Eqs.~\ref{eqVAnomega},\ref{eqVAprimenomega} to contain the $same$ unknown quantities, we had to choose $R=\vert Z(1\omega) \vert$ in order that $\left|V_{s,1}(1\omega)\right|=\left|V_{s,2}(1\omega)\right|$ because of the equilibration condition  (see Eq.~\ref{eqequilpont2s}). As a result Eqs~\ref{eqVAnomega} and \ref{eqVAprimenomega} can be easily solved.

Figure~\ref{figch1et2-88Hz} displays the harmonics $V_{s,1}(n\omega)$ of source 1 and $V_{s,2}(n\omega)$ of source $2$ obtained by using the two sources bridge method, for $\omega /(2\pi)$ = 88 Hz. The main features are: {\textit{i)} The second harmonics of both sources are similar in magnitude and phase. $\left|V_{s,i}(2\omega)\right|$ increases as the power two of the $1\omega$ voltage.{\textit{ii)} The third harmonics is twice smaller for source $2$ than for source $1$. One finds $\vert V_{s,2}(3\omega)\vert \propto \vert V_{s,2}(1\omega) \vert ^{2.6}$, i.e. a nearly cubic dependence which could have been mistaken with the physical signal if the tests mentionned above had not been performed. {\textit{iii)} The Total Harmonic Distortion (THD), given by summing up all the harmonics magnitudes $n \ge 2$ and dividing by the magnitude of the fundamental signal, is dominated by the second harmonics. Thus, from Fig.~\ref{figch1et2-88Hz}, at $V_s$ = 7 V the THD is of the order of $5 \times 10^{-5}$, well below the $10^{-4}$ specification which is usually ensured by high quality electronic devices. These features remain basically true for all the frequencies we studied in the 1-100 Hz range. Fig.~\ref{figch2-4Hz} gives the second and third harmonics of the second source at $4.28$ Hz that will be used in the next section. 

We summarize this section~\ref{Setup} by emphasizing the two main requirements that have to be met to measure the nonlinear response of our samples: First, the $1\omega$ part of the signal has to be suppressed before amplifying the sought $3\omega$ signal, and this can be done by using a bridge technique. Second the bridge must be balanced at $\omega$ \textit{and} $3\omega$.

\section{\label{New setup} A setup allowing the measurement of $\chi_3(\omega)$ of supercooled glycerol}

\subsection{\label{New setup1} One source bridge with two samples}

Let us consider the circuit depicted in the inset of Fig.~\ref{fig-gligli-n=2}. It uses only one source and is inspired from the Wheastone bridge, but it contains \textit{two liquid dielectric capacitors of different thicknesses $L_{thin}$ and $L_{thick}$}. The right arm of the bridge contains the thin capacitor (``thin sample'') of impedance $Z_{thin}\propto L_{thin}$ in series with a chosen impedance $z_{thin}$, while the left arm contains the thick capacitor (``thick sample'') of impedance $Z_{thick} \propto L_{thick}$ in series with a chosen impedance $z_{thick}$. The signal $V_m$ is measured between the two middle points of the two arms with our lock-in amplifier in differential mode. The input impedances of the lock-in are much larger than $\left|z_{thin}\right|$ and $\left|z_{thick}\right|$, thus the measured voltage ${V_{m,s}}$ at any frequency, due to the source voltage $V_s$ at the same frequency is
\begin{equation}
\frac{V_{m,s}(\omega)}{V_{s}(\omega)} = \frac{z_{thin}Z_{thick}-z_{thick}Z_{thin}}{(z_{thin}+Z_{thin})(z_{thick}+Z_{thick})},
\label{eqVmpont1s}
\end{equation}
where the impedances are taken at the frequency considered. The bridge is balanced for 
\begin{equation}
z_{thin}Z_{thick}=z_{thick}Z_{thin}.
\label{eqequilpont1s}
\end{equation}
The key point is that if we choose $z_{thin}$ and $z_{thick}$ of the same nature (e.g. two resistances, or two capacitances), once Eq.~\ref{eqequilpont1s} is fulfilled at a given frequency, \textit{it is fullfilled at any frequency}. This is because the relative variations of the two samples impedances with frequency are the same on both sides of Eq.~\ref{eqequilpont1s}. As a result, if the equilibration condition (Eq.~\ref{eqequilpont1s}) is fulfilled at $1 \omega$, all the harmonics $V_{s}(n\omega)$ generated by the source will give a zero contribution to $V_m$. This two-capacitors bridge thus allows to get rid of {\it both} the nonlinearities of the amplifier (see section~\ref{Setup b}) and of the harmonics generated by the source (see section~\ref{Setup c}). 

Obviously, the physical contributions from the samples yield a non-zero measured signal $V_{m,sample}(3\omega)$ because the two current sources at $3\omega$ corresponding to the samples verify $I_{thin}(3\omega)\propto L_{thin}^{-3}$ and $I_{thick}(3\omega)\propto L_{thick}^{-3}$ (see section~\ref{Setup a}). For a typical $L_{thick}$ $\approx$ $2L_{thin}$, the factor $8$ between the two currents is not compensated by the factor $2$ between $z_{thick}$ and $z_{thin}$: The relation between the measured signal and the two physical currents is 
\begin{equation}
V_{m,sample}(3\omega)  = \frac{I_{thin}(3\omega)}{y_{thin}(3\omega)  +  Y_{thin}(3\omega)} 
 -  \frac{I_{thick}(3\omega)}{y_{thick}(3\omega) + Y_{thick}(3\omega)} ,
\label{eqglycerolpont1s}
\end{equation}
where $Y_{thin}$ = $1/Z_{thin}$, $Y_{thick}$ = $1/Z_{thick}$, $y_{thin}$ = $1/z_{thin}$ and $y_{thick}$ = $1/z_{thick}$. By using Eqs~\ref{eqI3},\ref{eqequilpont1s} and $Y_{thick}/Y_{thin}=L_{thin}/L_{thick}$, Eq.~\ref{eqglycerolpont1s} becomes
\begin{equation}
V_{m,sample}(3\omega) = \frac{I_{thin}(3\omega)}{y_{thin}(3\omega) + Y_{thin}(3\omega)} \times\left(1-\left(\frac{L_{thin}}{L_{thick}}\right)^2\right).
\label{eqglycerolpont1sBis}¨
\end{equation}
In practice, $z_{thin}$ and $z_{thick}$ were resistors ($r_{thin}$ and $r_{thick}$), but we improved the equilibration by adding a small capacitor $c_{thin}$ in parallel with $r_{thin}$. $c_{thin}$ compensates for the unavoidable stray capacitances between the circuit and the ground. The equilibration was realized by tuning $r_{thin}$, $r_{thick}$ and $c_{thin}$ to minimize $\left|V_m\right|$ at $1\omega$. In practice, the equilibration at $n\omega$ ($n$ $>$ 1) is not perfect: Small asymetries between the two samples, stray capacitances which are not proportionnal to the samples thicknesses etc make that $V_m(n\omega)$ is not exactly zero. The bridge is thus characterized by the quality factors $\rho(n\omega) = \left|V_m(n\omega)/Vs(n\omega)\right|$, which can be measured by tuning the source fundamental frequency at $n\omega$. With setup A ($S$ $\simeq$ 5.5 cm$^2$, $L_{thin}$ $\simeq$ 30 $\mu$m, $L_{thick}$ $\simeq$ 60 $\mu$m, see section~\ref{Setup a}), we reached e.g. $\rho(1\omega) \simeq 3 \times 10^{-5}$, $\rho(2\omega) \simeq 5.9 \times 10^{-3}$ and $\rho(3\omega) \simeq 1.2 \times 10^{-2}$ for $\omega/2\pi$ = 27 Hz and $\omega \simeq \omega_{\alpha}$ ($T$ = 208 K). With setup B ($S$ $\simeq$ 3.14 cm$^2$, $L_{thin}$ $\simeq$ 19 $\mu$m, $L_{thick}$ $\simeq$ 41 $\mu$m), we obtained $\rho(1\omega) \simeq 2.7 \times 10^{-6}$, $\rho(2\omega) \simeq 6.6 \times 10^{-4}$ and $\rho(3\omega) \simeq 1.4 \times 10^{-3}$ for $\omega/2\pi$ = 29 Hz and $\omega \simeq \omega_{\alpha}$ ($T$ = 209.5 K). Note that smaller values of $\rho(3\omega)$ are obtained in the range $\omega \ll \omega_{\alpha}$ (i.e. when the capacitances of the samples are the largest): for example, with setup B, for $T=204.5K$ and $\omega / \omega_{\alpha} = 0.0567$, we find $\rho(3\omega) \simeq 2 \times 10^{-4}$ while $\rho(1\omega)$ is only one order of magnitude lower.

There is an optimum of the ratio $r_{thin}/\left|Z_{thin}(1\omega)\right|$ = $r_{thick}/\left|Z_{thick}(1\omega)\right|$ which maximizes the measured voltage $\left|V_{m,sample}(3\omega)\right|$. If this ratio decreases, the voltage $V_0$ = $V_s(1\omega)\left|Z_{thin}/(r_{thin}+Z_{thin})\right|$ applied to the samples increases, thus $\left|I(3\omega)\right|$ $\propto$ $V_0^3$ increases too; but at the same time, the contribution of $I(3\omega)$ to $V_{m,sample}(3\omega)$ decreases because it results from this current flowing through $r_{thin}$ (or $r_{thick}$) in parallel with $Z_{thin}(3\omega)$ (or $Z_{thick}(3\omega)$). We found that this optimum was $r_{thin}/\left|Z_{thin}(1\omega)\right|$ = $r_{thick}/\left|Z_{thick}(1\omega)\right|$ $\simeq$ 0.36. This value depends only weakly on the frequency and on $\tau_{\alpha}(T)$ (see section~\ref{Motivation}). 

\subsection{\label{results} Detection of the nonlinear dielectric response of glycerol}

We show in this section that by using the two sample bridge with setup A (see section~\ref{Setup a}) and the source 2 (whose harmonic distortion has been studied in section~\ref{Setup c}), we reached the sensitivity needed to detect the physical signal. We compare the measured voltage due to the source harmonics $V_{m,s}(3\omega)$ to the expected physical signal $V_{m,sample}(3\omega)$, for the maximum source voltage $V_{s,2}(\omega)$ = 7 V at $\omega/(2\pi)$ = 4.28 Hz ($\omega / \omega_{\alpha}$ $\approx$ 1 for $T$ = 203.7 K), with setup A ($L_{thin}$ = 30 $\mu$m, $L_{thick}$ = 60 $\mu$m). Using the results reported in Fig.~\ref{figch2-4Hz} and the measured $\rho(3\omega) \simeq 1.4\times 10^{-2}$, we find $\left|V_{m,s}(3\omega)/V_{s,2}(1\omega)\right| \approx 7\times10^{-8}$.
By using $\left|I(3\omega)/I(\omega)\right|$ $\simeq$ $1.1 \times 10^{-6}$ (see section~\ref{Setup a} and Eq~\ref{eqI3surI1}) and Eq.~\ref{eqglycerolpont1sBis}, we obtain $\left|V_{m,sample}(3\omega)/V_{s,2}(1\omega)\right| \approx 8 \times 10^{-8}$. The measured voltage is $V_m(3\omega)$ = $V_{m,s}(3\omega) + V_{m,sample}(3\omega)$, and the physical signal should contribute to $V_m(3\omega)$. 

Figs ~\ref{fig-gligli-n=2} and~\ref{fig-gligli-n=3} give the measured $V_m(2\omega)$ and $V_m(3\omega)$ for the same frequency 4.28 Hz and temperature 203.7 K. Fig.~\ref{fig-gligli-n=2} shows that the measured $V_m(2\omega)$ can be calculated by assuming that it is only due to $V_s(2\omega)$ generated by the source, in agreement with $I(2\omega)$ = 0 due to the field reversal symmetry mentionned in the introduction. On the contrary, Fig.~\ref{fig-gligli-n=3} shows a discrepancy between $V_m(3\omega)$ and the $V_{m,s}(3\omega)$ value that should be measured if it was only due to $V_s(3\omega)$ generated by the source, strongly suggesting that a physical signal is present. As for $n=2$, we used $\vert V_{m,s}(3\omega)\vert =\rho(3\omega) \vert V_s(3\omega)\vert$, with a proper treatment of the phase. The physical signal should grow as $V_s(\omega)^3$, while $V_{s,2}(3\omega)$ grows as $V_s(\omega)^{2.6}$ (see section ~\ref{Harmonics Source}). This difference of exponents explains that $V_{m}(3\omega)$ grows slower at high voltages where the physical signal should dominate progressively $V_{m,s}(3\omega)$. This also explains the evolution of the phase. The $V_{m}(3\omega)$ vs. $V_s(\omega)$ data can be fitted by a sum of $V_{m,s}(3\omega)$ and a physical signal $\lambda [V_{s}(\omega)]^3$ where $\lambda$ is a complex number. The same can be done at other temperatures and frequencies. However, we rather present in the next section the results obtained with setup B using another source: The improved quality factor $\rho$ and $\left|I(3\omega)/I(\omega)\right|$ ratio for this setup, and the lower THD of the source allowed the first accurate measurement of $\chi_3$. 

\section{\label{Final results} Measurement of the nonlinear dielectric response of a supercooled liquid}

We present in this section our first measurements of the nonlinear dielectric susceptibility of supercooled glycerol. We used the two samples bridge, with setup B (see section~\ref{Setup a}: $L_{thin} \simeq 19\ \mu$m, $L_{thick} \simeq 41\ \mu$m and $S$ = 3.14 cm$^2$). We used a voltage source with a lower total harmonic distortion (THD) than the sources used in sections~\ref{Setup},~\ref{New setup}: It was a SRS$^{\hbox{\textregistered}}$ DS360 ultra low distorsion function generator, with $V_s$ $\leq$ 14 V and a typical THD ranging from -98 dB to -109 dB (for $V_s$ = 10 V) in our frequency range. Note that due to the very low level of our physical signal, our bridge technique remains necessary even with such a low THD. Fig.~\ref{fig-gligli-final} shows the current $I_{thin}(3\omega)$ calculated from the measured $V_{m,sample}(3\omega)$ by using Eq.~\ref{eqglycerolpont1sBis}, as a function of the source voltage magnitude $\left|V_s\right|$, for $T$ = 210.2 K, $\omega/2\pi$ = $\omega_{\alpha}/2\pi$ = 43.76 Hz. A clear $\left|I_{thin}(3\omega)\right|\propto \left|V_{s}(\omega)\right|^3$ dependence is found, while the phase of $I(3\omega)$ does not depend on $\left|V_{s}(\omega)\right|$. These results are a strong indication that the origin of the signal is really the nonlinear response of the samples. A constant phase suggests that no other signal than the physical one is present, contrary to the results of the previous section. The magnitude is of the order of the estimated physical value $\left|I(3\omega) /I(\omega)\right|$ $\simeq$ $1.5\times 10^{-5}$ for $V_s$ = 14 V (see section~\ref{Setup}) which gives $\left|I(3\omega)\right|$ $\simeq$ $3\times 10^{-10}$ A.
 
To confirm the physical origin of our measurement, we compared these results to the data obtained by using a passive notch filter method. The principle of this method is depicted in the inset of Fig.~\ref{fig-gligli-final}: The source voltage $V_s(\omega)$ is applied to a sample capacitor (impedance $Z$) in series with a resistor $r$. The $3\omega$ voltage a point A ($V_A(3\omega)$) is measured by the lock-in amplifier. However, contrary to the method depicted in section~\ref{Setup b}, the fundamental component of the voltage at point $A$, $V_A(\omega)$ is strongly attenuated by a ``twin-T'' passive notch filter with a center frequency $f_c$ = $\omega/2\pi$. The twin-T notch filter is made of three capacitors $C$, $C$, 2$C$ and three resistors $2R$, $2R$ $R$, and $f_c$ = $1/2\pi RC$~\cite{electro}. A passive filter avoids the nonlinearities related to active components, and we verified that the resistors and capacitors did not yield noticeable $3\omega$ contribution to $V_A$. The attenuation coefficient at $1\omega$ (a few $10^{-4}$) was such that the $1\omega$ component at the lock-in input yielded a negligible $3\omega$ harmonics (see Fig.~\ref{figlockin}). To subtract the contribution to $V_A(3\omega)$ of the third harmonics due to the source $V_s(3\omega)$, the measurement was repeated with a circuit in which the sample was replaced by a resistor of impedance $\left|Z(\omega)\right|$. Finally, the currents $I(3\omega)$ obtained by using this method with the thin or the thick capacitor were proportional to the thicknesses at the power 3 as expected (see Eq.~\ref{eqI3}). The current $I(3\omega)$ obtained with the twin-T filter method is in remarkable agreement with the current obtained using the two sample bridge, both in what concerns the phase and the magnitude (see Fig. \ref{fig-gligli-final}).

We also confirmed the purely physical origin of our results of Fig.~\ref{fig-gligli-final} by verifying that the two possible contributions to $V_m(3\omega)$ due to the source and to the lock-in amplifier were negligible in the case of the two samples bridge. As explained above, $V_s(3\omega)$ was measured by using the twin-T notch filter method with the sample replaced by a resistor. Its value depends on $\omega$ and $V_s$, but we found that its contribution to $V_m(3\omega)$ remained much smaller than the physical contribution for the two samples bridge, e.g. for $T$ = 210.2 K and $\omega/(2\pi)$ = 43.76 Hz, it is 3 orders of magnitude below. The contribution of the lockin was negligible because the equilibration of the bridge led to $V_m(\omega)$ values below 1 mV: Fig.~\ref{figlockin} shows that the resulting third harmonics is well below 40 nV, a negligible value with respect to the measured $V_m(3\omega)$ of a few $\mu$V. This remained true when the lockin-in amplifier was used in the differential mode provided that the voltages on each input remained below about 1.4 V. Another possible spurious contribution to the measured third harmonics could come from the resistors $r_{thin}$, $r_{thick}$ and the capacitor $c_{thin}$. We verified in experiments using bridges made only of resistors or of capacitors or of combinations of both that the nonlinearities of metal film resistors and polystyrene film capacitors (we rejected ceramic capacitors) were negligible. Finally, another verification of the physical origin of our data was that $\left|V_m(2\omega)\right|$ remained much weaker than $\left|V_m(3\omega)\right|$.

\section{\label{Extensions} Possible developments of the two samples bridge method}

The equilibration (Eq.~\ref{eqequilpont1s}) of the two-sample bridge is realized by tuning $z_{thin}$ and $z_{thick}$ (two resistors in our case) in series with the two impedances of the samples, $Z_{thin}$ and $Z_{thick}$ (see section~\ref{New setup} and the insets of Figs.~\ref{fig-gligli-n=2},\ref{fig-gligli-n=3}). This is time consuming because in order to keep the ratio $r_{thin}/\left|Z_{thin}(1\omega)\right|$ = $r_{thick}/\left|Z_{thick}(1\omega)\right|$ at its optimum value, it has to be done for each temperature or frequency change. A possible solution is a {\it four} samples bridge: In such a device, the two sample capacitors of different thicknesses are kept, but the two impedances $z_{thin}$ and $z_{thick}$ are replaced by two other sample capacitors. Thus, neither the equilibration, nor the ratio of impedances in each arm depend on $T$ or $\omega$.
The first (resp. second) arm of a four sample bridge would contain sample 1 (resp. 2) connected to the source and sample 3 (resp. 4) connected to the ground. The thicknesses are $L_i$ ($i$ = 1,4) and the electrode surfaces $S_i$ ($i$ = 1,4). If the surfaces are the same in an arm ($S_1$ = $S_3$ or $S_2$ = $S_4$), the contribution of the two physical currents ($I_1(3\omega)$ and $I_3(3\omega)$, or $I_2(3\omega)$ and $I_4(3\omega)$) to the measured $3\omega$ voltage $V_{m,sample}(3\omega)$ is zero. This is because the electric field at $1\omega$ in the two capacitors in an arm would be the same: The different thicknesses (see Eq.~\ref{eqI3}) are compensated by the voltage division between the two capacitors. Thus $S_1$ $\neq$ $S_3$ and $S_2$ $\neq$ $S_4$ are needed.  The balancing condition $Z_1 Z_4$ = $Z_2 Z_3$ leads to $L_1 L_4 S_2 S_3$ = $L_2 L_3 S_1 S_4$. Following the lines of section~\ref{New setup1}, the measured $3\omega$ voltage is given by
\begin{equation}               
\frac{V_{m,sample}(3\omega)}{V_s^3(1 \omega)} \propto \frac{S_1S_3L_1L_3}{(L_1S_3 + L_3S_1)^4}(S_{3}^2 - S_{1}^2)-
\frac{S_2S_4L_2L_4}{(L_2S_4 + L_4S_2)^4}(S_{4}^2 - S_{2}^2).
\label{eq4samples} 
\end{equation}    

The two (or four) sample capacitors bridge methods can be generalized to any physical situation in which a sinusoidal excitation $F(t)$ = $Re[F_0e^{i\omega t}]$ gives a response $R(t)$ = $Re[\chi_1(\omega)F_0e^{i\omega t}$ + $\frac{1}{4} \chi_3(\omega)F_0^3e^{3i\omega t} + ...]$ where $\chi_3$, $\chi_5$ are to be measured. If $F_0$ depends on a parameter $\Lambda$ analogous to our capacitor thickness, it is possible to design a bridge in which each arm contains an experimental unit with two different values of $\Lambda$. Again, the advantage of such a bridge with respect to a bridge where the sample response is balanced by the response of an impedance~\cite{Gul65,Ros65,Jun92,Lee93,Bir97,Gar01} (or possibly with two sources as discussed in section~\ref{Setup c}) is that if $\chi_1$ depends on the frequency, the equilibration at $1\omega$ implies also an equilibration at $n\omega$. Among the possible applications of this principle, we can think of a bridge devoted to the measurement of the non linear magnetic susceptibility: The capacitors of our bridge would be replaced by coils with sample cores, and the $\Lambda$ parameter would be the number of turns of the coils.   

\section{Conclusion}

Many experiments in physics consit in applying an ``excitation'' (electric, magnetic, mechanical, etc.) to a sample at a finite frequency and measuring the response. The nonlinear part of this response is present in its harmonics. Their measurement is a hard task for weak nonlinear responses because the electronic devices used for the excitation and for the measurement are always non linear at a certain level.
Our experiment, devoted to the measurement of the nonlinear response of dielectric supercooled liquids, measures the third harmonics component of the polarization when an electric field is applied to a sample layer in a plane capacitor. It consists in measuring a current $I(3\omega)$ at a frequency three (possibly five, etc.) times that of the linear response $I(\omega)$ when a voltage $V_s(\omega)$ is applied to the capacitor. Both the lock-in amplifier and the voltage source yield a third harmonics component which precludes measuring the nonlinear response in a simple experiment such as this presented in section~\ref{Setup b}. To overcome this problem, we have used a method based on a bridge with two capacitors of different thicknesses. We have shown that such a device strongly reduces the two spurious third harmonics component of the measured signal: That of the voltage source and that of the lock-in amplifier. This ``two-sample'' bridge, used with a low distortion voltage source allowed to reach our goal of a sensitivity better than 10$^{-7}$ and to realize our first measurements of the nonlinear response of a supercooled liquid close to the glass transition.

\begin{acknowledgments}
We wish to thank Patrick Pari and Philippe Forget of the SPEC Cryogenics Laboratory for their valuable help in designing and realizing the cryostat and the experimental cell. This work was supported in part by A.N.R. under Grant "DynHet".
\end{acknowledgments}

\newpage 



\begin{figure*}
\includegraphics[scale=0.350,angle=0]{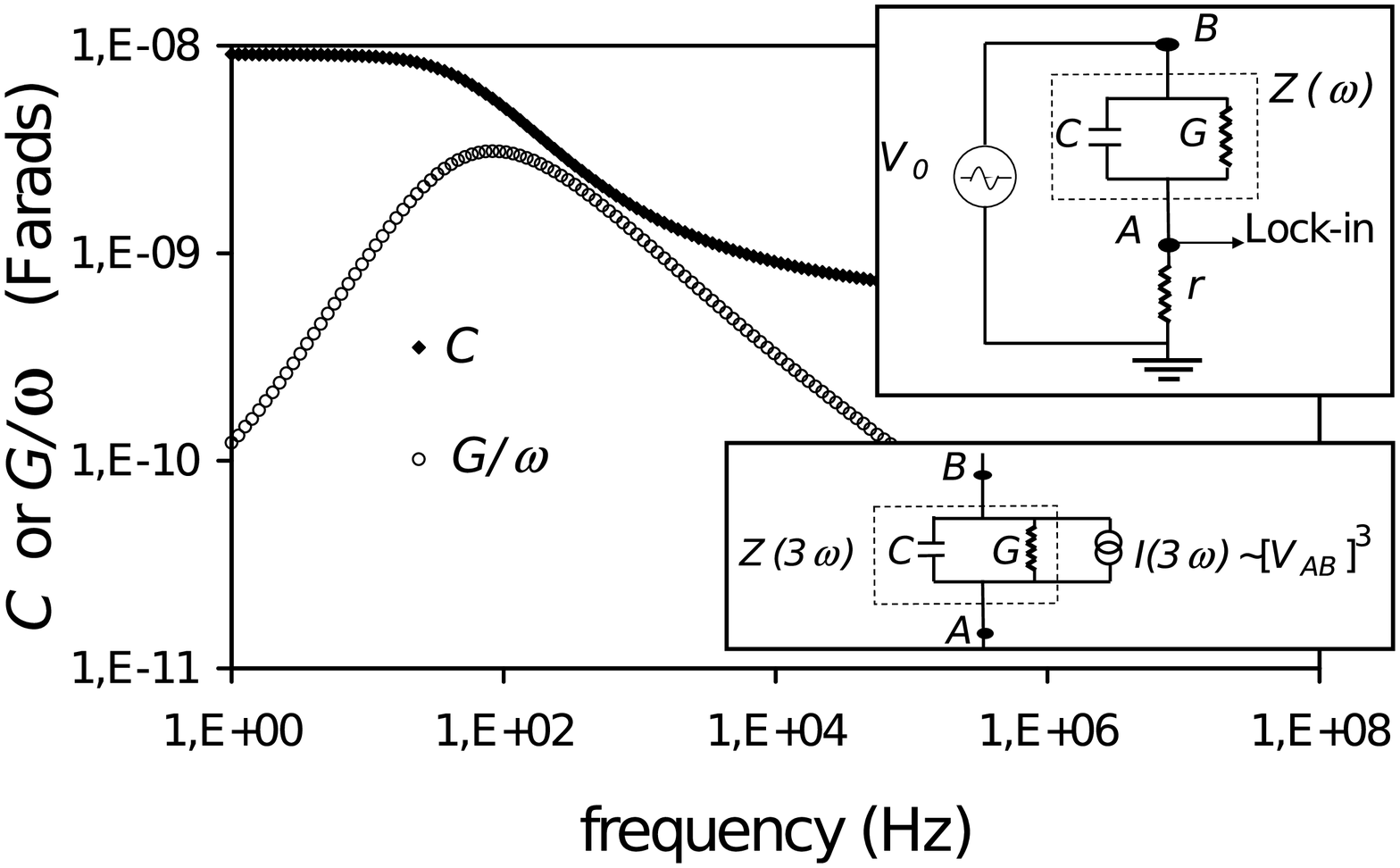}
\caption{\label{figchideomega} Measurement of the dielectric susceptibility $\epsilon$ = $\epsilon' + i\epsilon''$ = $\chi_1 + 1$. The experiment is shown in the upper inset: The capacitor with glycerol as dielectric material is equivalent to a non-dissipative capacitor $C(\omega)$ $\propto$ $\epsilon'$ in parallel with a conductance $G(\omega)$ $\propto$ $\omega \epsilon''$. The impedance $Z(\omega)$ = $(G + iC\omega)^{-1}$ is obtained by measuring the voltage drop across the resistor $r$ at point $A$ when the excitation voltage $V$ = $V_s \cos\omega t$ is applied in $B$. 
The measured values of $C$ and $G/\omega$ are shown as a function of $\omega/2\pi$ for $T$ = 211.8 K. In the lower right corner, the equivalent circuit corresponding to the physical third harmonics source is shown: an ideal current source in parallel with a capacitor $C(3\omega)$ and a conductance $G(3\omega)$.}
\end{figure*}

\begin{figure*}
\includegraphics[scale=0.350,angle=0]{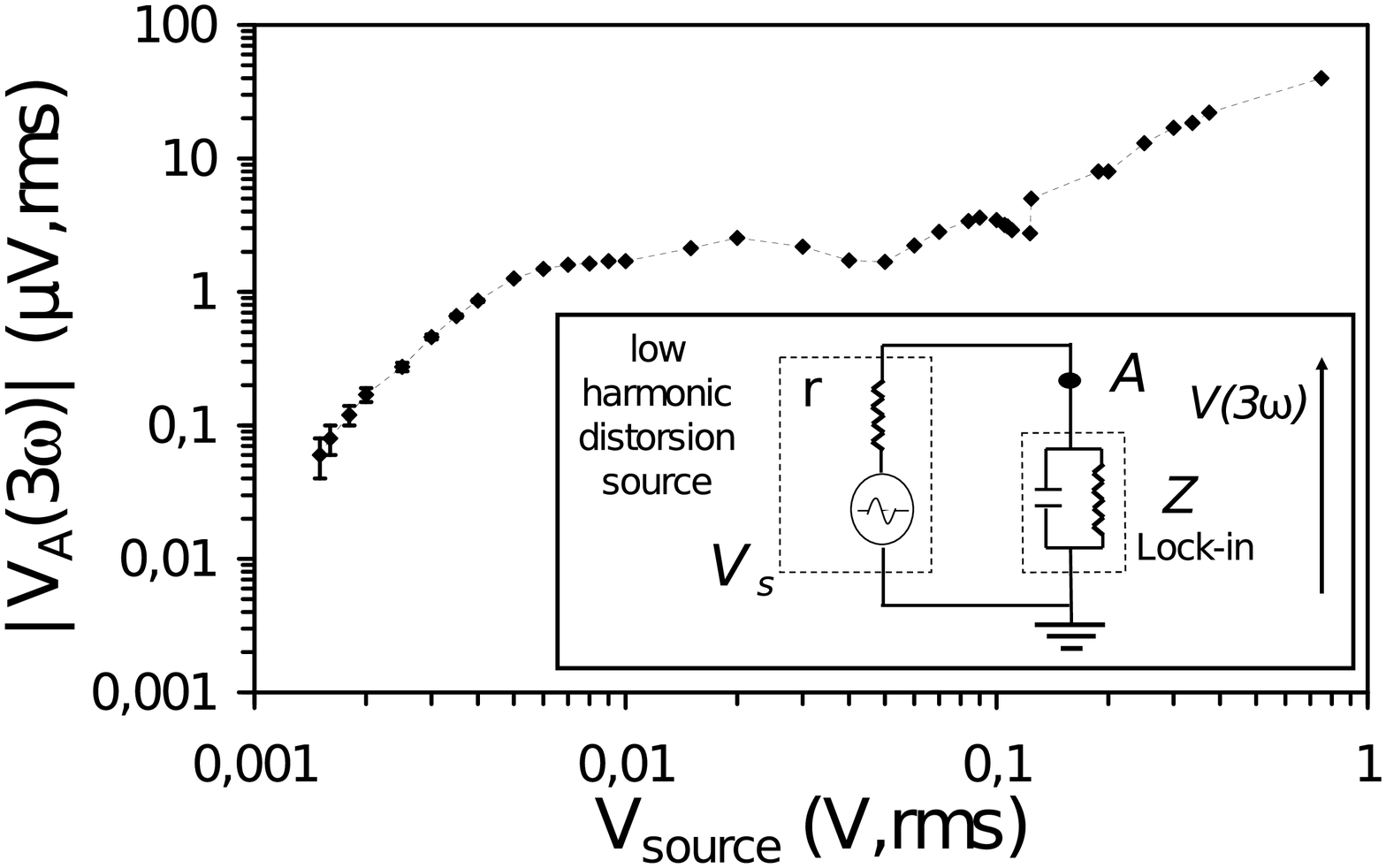}
\caption{\label{figlockin} Magnitude of the measured third harmonics voltage as a function of the source voltage when the latter is directly applied at the input of the lock-in amplifier (for $\omega/(2\pi)$ = 88 Hz).} 
\end{figure*}

\begin{figure*}
\includegraphics[scale=0.350,angle=0]{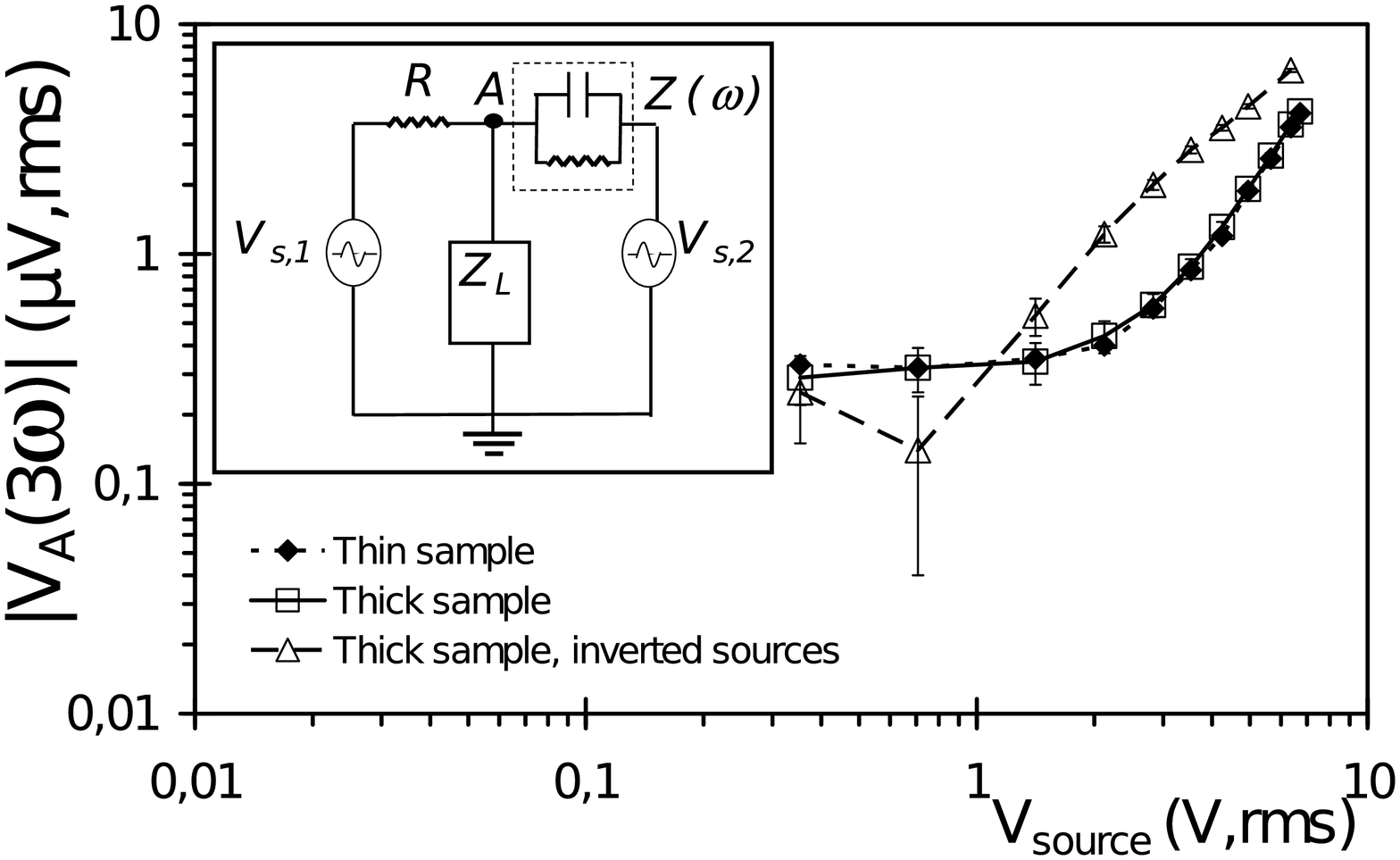}
\caption{\label{figpont2sources} The third harmonics voltage $\left|V_A(3\omega)\right|$ measured at point $A$ vs. the source voltage $\left|V_{s}(\omega)\right|=max(\left|V_{s,1}(1\omega)\right|,\left|V_{s,2}(1\omega)\right|)$ applied to the sample for the two sources bridge depicted in the inset. The components of the circuit are the sample (impedance $Z$), the two voltage sources ($V_{s,1}(\omega)$ and $V_{s,2}(\omega)$), the resistor $R$ and the lock-in input impedance $Z_L$. Closed diamonds with short dashed line: Source 2 applied to a thin sample ($L_{thin}$ = 30 $\mu$m). Open squares with continuous line:  Same circuit, but with a thick sample ($L_{thick}$ = 60 $\mu$m). Open triangles with long dashed line: Source 1 is applied to the thin sample, and source 2 to $R$. The temperature of the samples is 211.8 K and $\omega/(2\pi)$ = $\omega_{\alpha}/(2\pi)$ = 88 Hz}
\end{figure*}

\begin{figure*}
\includegraphics[scale=0.350,angle=0]{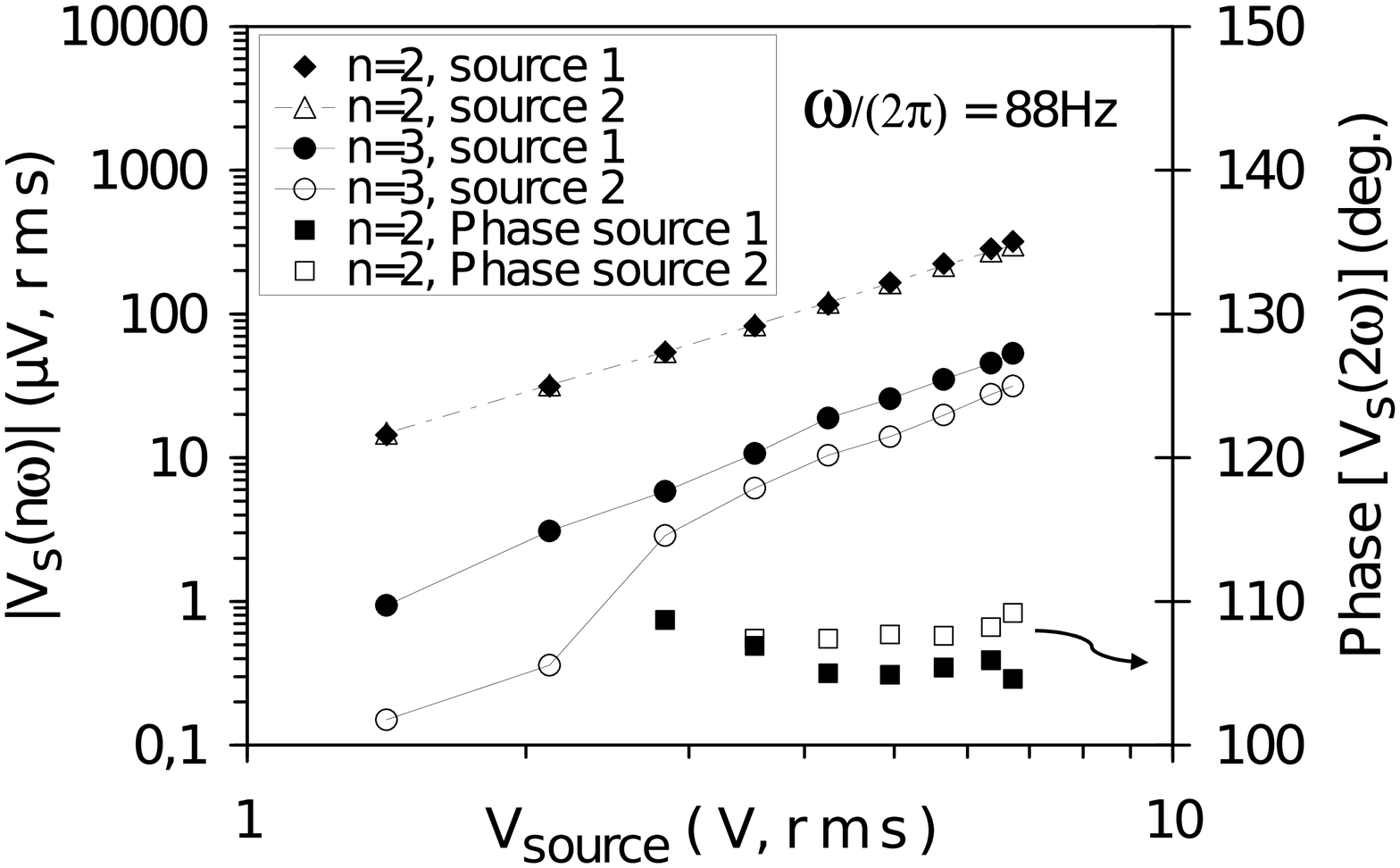}
\caption{\label{figch1et2-88Hz} Magnitude (left axis) and phase (right axis) of the second and third harmonics generated by the source 1 (closed symbols) and source 2 (open symbols) of the two sources bridge, obtained by solving Eqs.~\ref{eqVAnomega},\ref{eqVAprimenomega} for a frequency of 88 Hz.} 
\end{figure*}

\begin{figure*}
\includegraphics[scale=0.350,angle=0]{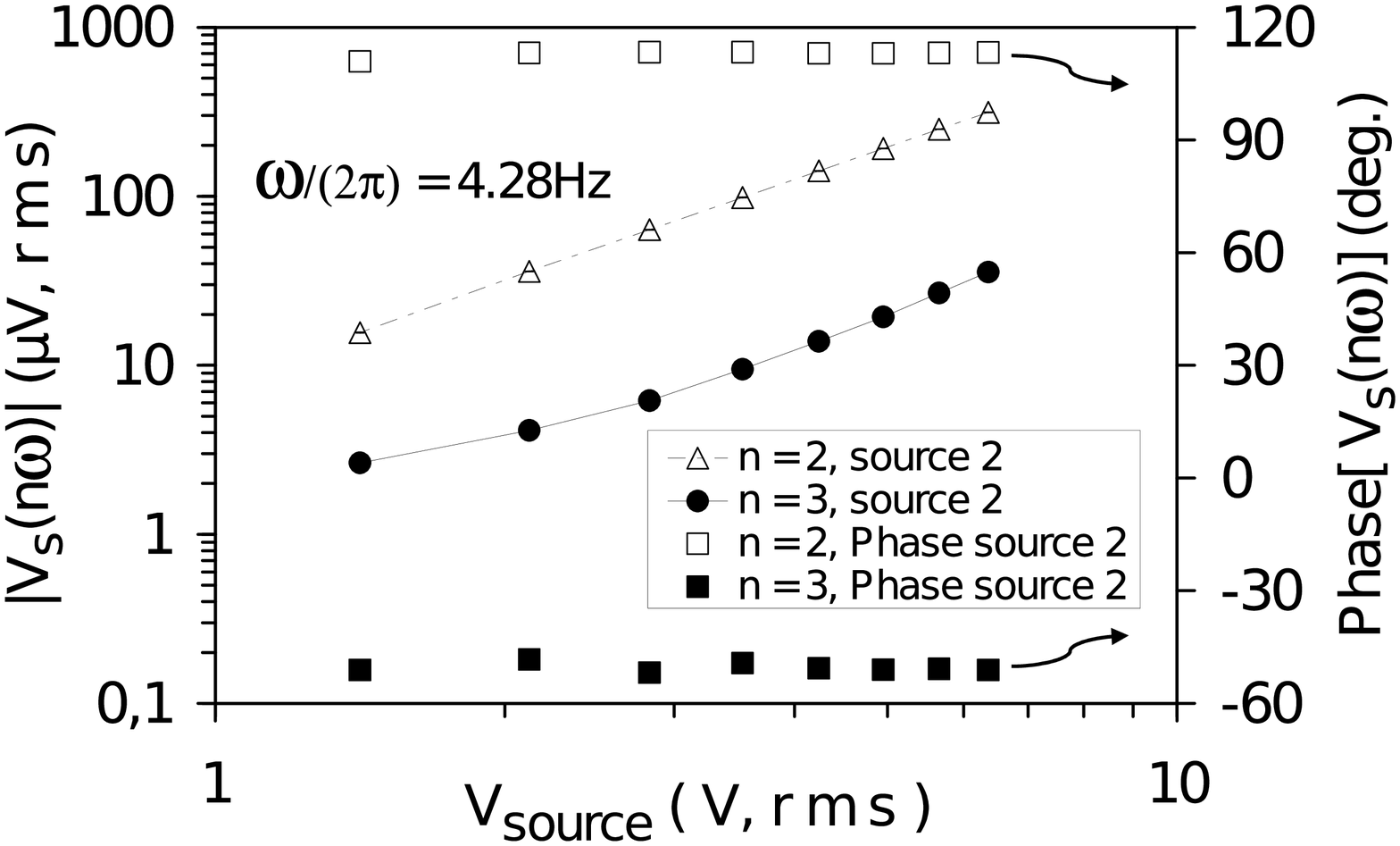}
\caption{\label{figch2-4Hz}Magnitude (left axis) and phase (right axis) of the second and third harmonics generated by the source 2 of the two sources bridge, obtained by solving Eqs.~\ref{eqVAnomega},\ref{eqVAprimenomega} for a frequency of 4.28 Hz.}
\end{figure*}

\begin{figure*}
\includegraphics[scale=0.37,angle=-90]{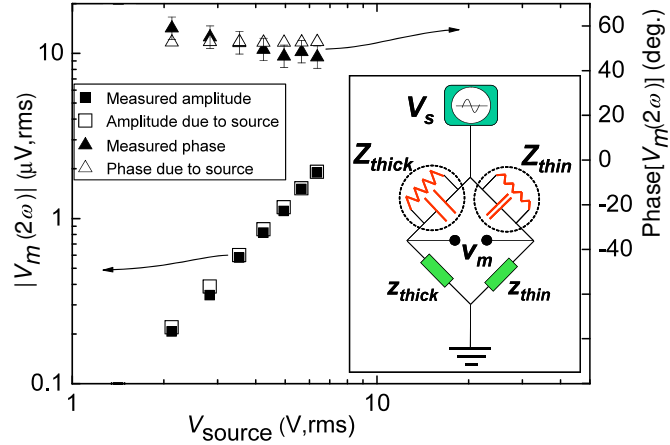}
\caption{\label{fig-gligli-n=2} Magnitude (left axis) and phase (right axis) of the second harmonics $V_m(2\omega)$ measured in the two samples bridge depicted in the inset. The experimental data (closed symbols) are compared to the values (open symbols) calculated by assuming that $V_m(2\omega)$ is only due to the second harmonics generated by the source $V_s(2\omega)$ which was obtained from the two sources measurements: $\left|V_m(2\omega)\right|$ = $\rho(2\omega)\left|V_s(2\omega)\right|$. $\omega/(2\pi)$ = 4.28 Hz and $T$ = 203.7 K}
\end{figure*}

\begin{figure*}
\includegraphics[scale=0.35,angle=-90]{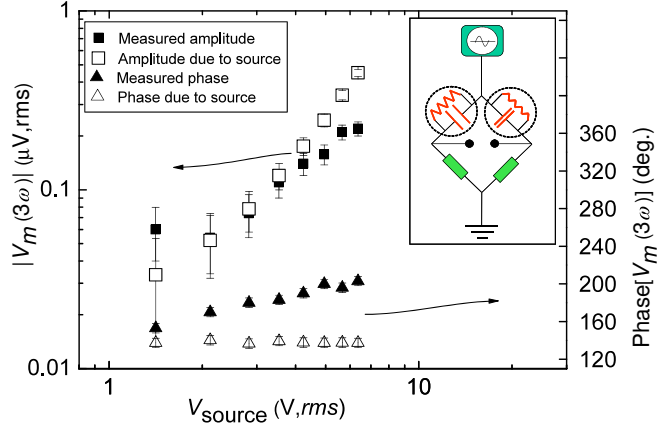}
\caption{\label{fig-gligli-n=3}Magnitude (left axis) and phase (right axis) of the third harmonics $V_m(3\omega)$ measured in the two samples bridge depicted in the inset. The experimental data (closed symbols) are compared to the values (open symbols) calculated by assuming that $V_m(3\omega)$ would be only due to the third harmonics generated by the source $V_s(3\omega$). $\omega/(2\pi)$ = 4.28 Hz and $T$ = 203.7 K}
\end{figure*}

\begin{figure*} 
\includegraphics[scale=0.39,angle=-90]{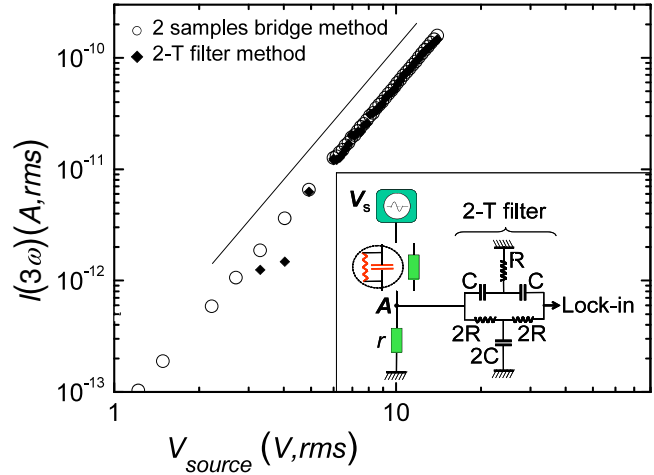}
\caption{\label{fig-gligli-final} Open circles: The current $\left|I_{thin}(3\omega)\right|$ calculated from the measured $V_{m,sample}(3\omega)$ in the two samples bridge, as a function of the source voltage magnitude $\left|V_s(\omega)\right|$, for $T$ = 210.2 K, $\omega/2\pi$ = 43.76 Hz. Closed diamonds: $\left|I_{thin}(3\omega)\right|$ obtained using twin-T filter method. The continuous line is the power 3 dependence. Inset: The twin-T notch filter method (see text).  }
\end{figure*}


\end{document}